\documentclass[10pt]{article}

\usepackage{amsmath,amsthm}

\newtheorem{thm}{Theorem}[section]

\newtheorem{prop}{Proposition}
\newtheorem{lem}[thm]{Lemma}
\def\n{\noindent}
\theoremstyle{remark}
\newtheorem*{rmk}{Remark}

\theoremstyle{plain}

\newtheoremstyle{note}
  {3pt}
  {3pt}
  {}
  {}
  {\itshape}
  {:}
  {.5em}
  {}

\theoremstyle{note}

\newtheoremstyle{citing}
  {3pt}
  {3pt}
  {\itshape}
  {}
  {\bfseries}
  {.}
  {.5em}
  {\thmnote{#3}}

\theoremstyle{citing}

\newtheoremstyle{break}
  {9pt}
  {9pt}
  {\itshape}
  {}
  {\bfseries}
  {.}
  {\newline}
  {}

\theoremstyle{break}

\swapnumbers
\theoremstyle{plain}

\let\lvert=|\let\rvert=|

\begin{document}

\begin{center} 
{\Large \bf On the Dressing Method for the Generalised \vskip5pt Zakharov-Shabat System}

\vskip.8cm
 
{\Large \it Rossen Ivanov\footnote{E-mail: Rossen.Ivanov@dcu.ie,\, \,  Tel:  + 353--1--700 5856, \, \,
Fax:  + 353--1--700 5508}}  
\vskip.8cm
School of Electronic Engineering, Dublin City University, Glasnevin, Dublin 9, Ireland\footnote{address for correspondence} \\
and Institute for Nuclear Research and Nuclear Energy, 72 Tzarigradsko chaussee, 1784 Sofia, Bulgaria 

\end{center}

\vskip1.32cm

\begin{abstract}
\noindent
The dressing procedure for the Generalised Zakharov-Shabat system is
well known for systems, related to $sl(N)$ algebras. We extend the method,
constructing explicitly the dressing factors for some systems,
related to orthogonal and symplectic Lie algebras. We consider 'dressed'
fundamental analytical solutions with simple poles at the prescribed
eigenvalue points and obtain the corresponding Lax potentials, representing
the soliton solutions for some important nonlinear evolution equations.

\vskip.4cm 


\vskip.4cm 

\end{abstract}

\section{Introduction}

The Non-linear evolution equations (NLEE), solvable by the inverse
scattering method (ISM) can be represented as a compatibility condition of
two linear systems with spectral parameter $\lambda $: 
\begin{equation}
\lbrack L(\lambda ),M(\lambda )]=0  \label{compatibility}
\end{equation}

\n
The ISM is based on the fact that the scattering data for the corresponding
equations satisfy linear equations, which are trivially solved \cite
{ZMNP,ZM,ZM1,K,FaTa,KRB}. One of the simplest, but at the same time the most
important systems for ISM is the so called Zakharov-Shabat (ZS) system \cite
{ZMNP}, \cite{AKNS}.

\begin{eqnarray}
L\psi _{0}(x,t,\lambda ) &\equiv &\left( i\frac{d}{dx}+q_{0}(x,t)-\lambda
\sigma _{3}\right) \psi _{0}(x,t,\lambda )=0, \label{su(n) ZS} \\
q_{0}(x,t)&=&q^{+}\sigma_{+}+q^{-}\sigma _{-}   \nonumber\\
\sigma _{+} &=&\left( 
\begin{array}{cc}
0 & 1 \\ 
0 & 0
\end{array}
\right) ,\;\sigma _{-}=\left( 
\begin{array}{cc}
0 & 0 \\ 
1 & 0
\end{array}
\right) ,\;\sigma _{3}=\left( 
\begin{array}{cc}
1 & 0 \\ 
0 & -1
\end{array}
\right)   \nonumber
\end{eqnarray}
The class of NLEE related to Lax operators of the form (\ref{su(n) ZS})
contains physically important equations such as the non-linear Schroedinger
equation (NLS), the sine-Gordon and MKDV equations and so on. The $N$-wave
equation requires $3\times 3$ matrix-valued potential $q_{0}(x,t)$ (see
bellow). For example, if the $M$-operator has the form 
\[
M\psi _{0}(x,t,\lambda )\equiv \left( i\frac{d}{dt}+V_{0}(x,t)+\lambda
V_{1}(x,t)-\lambda ^{2}\sigma _{3}\right) \psi _{0}(x,t,\lambda )
\]
and if we enforce the reduction $q\equiv q^{+}=-\overline{(q^{-})}$ then the
compatibility condition (\ref{compatibility}) can be resolved for $q$, $V_{0}
$, $V_{1}$ to give the NLS equation 
\[
iq_{t}+q_{xx}+|q|^{2}q(x,t)=0
\]
This construction clearly corresponds to a potential $q_{0}\in su(2)$ and
can be extended for all (semi)simple Lie algebras, leading to the so called
Generalised ZS system.

\n
In Section 2 we briefly describe the generalised ZS system and the ZS
dressing method. The ZS dressing method \cite{ZMNP}, \cite{Sh}, \cite{Za*Sh}
leads to a construction of a new solution, starting from a known one. The
dressing procedure like the Backlund transform and Darboux transform creates
a Lax operator, which has a new pair of complex discrete eigenvalues at
prescribed positions with respect to the original Lax operator. This method
is well known for systems, related to $sl(N)$ algebras \cite{ZMNP}, \cite
{Za*Sh}, \cite{VSG PPK}, \cite{TerUl}. The discreet spectral values of $L$
are related to the 'reflectionless' potentials of $L$ and soliton solutions
of the NLEE. In Section 3 we extend the method, constructing explicitly the
dressing factors and the new solutions for some important cases, related to
orthogonal and symplectic Lie algebras. We consider 'dressed' fundamental
analytical solutions with simple poles at the prescribed eigenvalue points
and obtain the corresponding Lax potentials, representing the soliton
solutions for some important NLEE such as N-wave and NLS-type equations.

\section{Generalised Zakharov-Shabat System}

\subsection{Properties of the Generalised Zakharov-Shabat System}

The Lax operator of the Generalised Zakharov-Shabat System has the form

\begin{equation}
L\psi (x,\lambda )\equiv \left( i\frac{d}{dx}+q(x)-\lambda J\right) \psi
(x,\lambda )=0  \label{ZS}
\end{equation}
where $q(x)$ and $J$ take values in the simple Lie-algebra $\bf{g}$ with a
Cartan subalgebra $\bf{h}$: $q(x)\in \bf{g}\mathbf{/}\bf{h}$ is a
Schwartz-type function i.e. vanishing fast enough for $|x|\to \infty $ , $%
J\in \bf{h}$ is a real constant regular element. The regularity of $J$
means that $\alpha (J)>0$ for all positive roots $\alpha \in \Delta _{+}$ of 
$\bf{g}$. The continuous spectrum of $L$ (\ref{ZS}) fills up the real line 
$\bf{R}$ in the complex $\lambda $-plane. \\
Here we fix up the notations and the normalization conditions for the
Cartan-Weyl generators of $\bf{g}$. The commutation relations are given by 
\cite{LA}: 
\begin{eqnarray}
&&[H_{e_{k}},E_{\alpha }]=(\alpha ,e_{k})E_{\alpha },\quad [E_{\alpha
},E_{-\alpha }]=H_{\alpha },  \nonumber \\
&&[E_{\alpha },E_{\beta }]=\left\{ 
\begin{array}{ll}
N_{\alpha ,\beta }E_{\alpha +\beta }\quad & \;\alpha +\beta \in \Delta \\ 
0 & \;\alpha +\beta \notin\Delta \cup \{0\}.
\end{array}
\right.
\end{eqnarray}
where $\Delta $ is the root system of $\bf{g}$ , $H_{e_{k}}$, $%
k=1,...,r$ are the Cartan subalgebra generators and $E_{\alpha }$ are the
root vectors of the simple Lie algebra $\bf{g}$. Here and below $r=\mathrm{%
rank}(\bf{g}\mathbf{)}$, and $e_{k}$, $\alpha $, $\vec{J}%
=\sum_{k=1}^{r}J_{k}e_{k}$ $\in {\bf{E}}^{r}$ are Euclidean vectors
corresponding to the Cartan elements $H_{e_{k}}$, $H_{\alpha }$ and $J$ $%
=\sum_{k=1}^{r}J_{k}H_{e_{k}}$ correspondingly. The normalization of the
basis is determined by: 
\begin{eqnarray}
&&E_{-\alpha }=E_{\alpha }^{T},\quad \mathrm{tr}(E_{-\alpha }E_{\alpha })={%
\frac{(\alpha ,\alpha )}{2}},  \nonumber \\
&&N_{-\alpha ,-\beta }=-N_{\alpha ,\beta },\quad N_{\alpha ,\beta }=\pm
(p+1),
\end{eqnarray}
where the integer $p\geq 0$ is such that $\alpha +s\beta \in \Delta $ for
all $s=1,\dots ,p$ and \linebreak $\alpha +(p+1)\beta \notin\Delta $. \\
We can define fundamental analytic solutions (FAS) $\chi ^{\pm }(x,\lambda )$
of $L$, which are analytic functions of $\lambda $ for $\pm \mathrm{Im}\,\lambda
> 0$ as follows:

\begin{eqnarray}
\lim_{x\to -\infty }e^{i\lambda Jx}\chi ^{\pm }(x,\lambda ) &=&S^{\pm
}(\lambda )\text{,\ }  \nonumber \\
\lim_{x\to \infty }e^{i\lambda Jx}\chi ^{\pm }(x,\lambda ) &=&T^{\mp
}(\lambda )D^{\pm }(\lambda )  \label{FAS}
\end{eqnarray}
where $S^{\pm }(\lambda )$, $D^{\pm }(\lambda )$ and $T^{\pm }(\lambda )$
are the factors in the Gauss decomposition of the scattering matrix $%
T(\lambda )$ \cite{ZMNP,Sh}:

\begin{equation}
T(\lambda )=T^{-}(\lambda )D^{+}(\lambda )\hat{S}^{+}(\lambda
)=T^{+}(\lambda )D^{-}(\lambda )\hat{S}^{-}(\lambda ).  \label{Gaussdecomp}
\end{equation}
Here by ''hat'' above we denote the inverse matrix $\widehat{S}\equiv S^{-1}$%
. It is convenient to use the following parametrization for the factors in (%
\ref{Gaussdecomp})

\begin{eqnarray}
&&S^{\pm }(\lambda )=\exp \left( \sum_{\alpha \in \Delta _{+}}s_{\alpha
}^{\pm }(\lambda )E_{\pm \alpha }\right) ,\quad T^{\pm }(\lambda )=\exp
\left( \sum_{\alpha \in \Delta _{+}}t_{\alpha }^{\pm }(\lambda )E_{\pm
\alpha }\right) ,  \nonumber \\
&&D^{+}(\lambda )=\exp \left( \sum_{j=1}^{r}{\frac{2d_{j}^{+}(\lambda )}{%
(\alpha _{j},\alpha _{j})}}H_{j}\right) ,\quad D^{-}(\lambda )=\exp \left(
\sum_{j=1}^{r}{\frac{2d_{j}^{-}(\lambda )}{(\alpha _{j},\alpha _{j})}}%
H_{j}^{-}\right) ,  \label{GaussTSD}
\end{eqnarray}
where $H_{j}\equiv H_{\alpha _{j}}$,  $\{\alpha _{j}\}_{j=1}^{r}$ is the set of the 
simple roots of  $\bf{g}$, $H_{j}^{-}=w_{0}(H_{j})$ and $w_{0}$ is
the Weyl group element which maps the highest weight of each irreducible
representation to the corresponding lowest weight. The proof of the
analyticity of $\chi ^{\pm }(x,\lambda )$ for any semi-simple Lie algebra
and real $J$ is given in \cite{VG*86} (for $sl(N)$ in \cite{ZMNP,ZM,ZM1,Sh}%
). The upper scripts $+$ and $-$ in $D^{\pm }(\lambda )$ show that $%
D_{j}^{+}(\lambda )$ and $D_{j}^{-}(\lambda )$: 
\begin{equation}
D_{j}^{\pm }(\lambda )=\langle \omega _{j}^{\pm }|D^{\pm }(\lambda )|\omega
_{j}^{\pm }\rangle =\exp \left( d_{j}^{\pm }(\lambda )\right) ,\qquad \omega
_{j}^{-}=w_{0}(\omega _{j}^{+}),  \label{d_j}
\end{equation}
are analytic functions of $\lambda $ for $\mathrm{Im}\,\lambda >0$ and $%
\mathrm{Im}\,\lambda <0$ respectively. Here $\omega _{j}^{+}$ are the
fundamental weights of $\bf{g}$ and $|\omega _{j}^{+}\rangle $ and $%
|\omega _{j}^{-}\rangle $ are the highest and lowest weight vectors in these
representations. On the real axis $\chi ^{+}(x,\lambda )$ and $\chi
^{-}(x,\lambda )$ are related by

\begin{equation}
\chi ^{+}(x,\lambda )=\chi ^{-}(x,\lambda )G_{0}(\lambda ),\qquad
G_{0}(\lambda )=S^{+}(\lambda )\hat{S}^{-}(\lambda ),  \label{RH probl}
\end{equation}
and the sewing function $G_{0}(\lambda )$ may be considered as a minimal set
of scattering data provided the Lax operator (\ref{ZS}) has no discrete
eigenvalues. The presence of discrete eigenvalues $\lambda _{1}^{\pm }$
means that one (or more) of the functions $D_{j}^{\pm }(\lambda )$ will have
zeroes at $\lambda _{1}^{\pm }$ ($\mathrm{Im}\,\lambda _{1}^{+}>0$, $\mathrm{%
Im}\,\lambda _{1}^{-}<0$ ). The Riemann-Hilbert (RH) problem (\ref{RH probl}%
) is equivalent to the system (\ref{ZS}) for $\chi ^{\pm }$.

\subsection{Non-linear evolution equations}

We can introduce a dependence on an additional 'time' parameter $t$. The
non-linear evolution equations possess a Lax representation of the form (\ref
{compatibility}) where 
\begin{eqnarray*}
M(\lambda )\psi (x,t,\lambda ) &\equiv &\left( i{\frac{d}{dt}}+V(x,t,\lambda
)\right) \psi (x,t,\lambda )=0,\quad \\
V(x,t,\lambda ) &=&\sum_{k=0}^{P-1}V_{k}(x,t)\lambda ^{k}-f_{P}\lambda
^{P}I,\quad V_{k}\in {\bf{g}},\;I\in {\bf h}\text{, }I=\mathrm{const}
\end{eqnarray*}
which must hold identically with respect to $\lambda $. The components of $I$
are real also. A standard procedure generalizing the AKNS one \cite{AKNS}
allows us to evaluate $V_{k}$ in terms of $q(x,t)$ and its $x$-derivatives.
Here and below we consider only the class of Schwartz-type potentials $%
q(x,t) $ vanishing fast enough for $|x|\to \infty $ for any fixed value of $%
t $. Then one may also check that the asymptotic value of the potential in $%
M $, namely $f^{(P)}(\lambda )=f_{P}\lambda ^{P}I$ may be understood as the
dispersion law of the corresponding NLEE. \\
For example, the $N$--wave equation \cite{ZMNP,ZM,ZM1,K,KRB} 
\begin{equation}
i[J,Q_{t}]-i[I,Q_{x}]+[[I,Q],[J,Q]]=0,  \label{N wave}
\end{equation}
corresponds to a generalized ZS type system with $q(x,t)\equiv [J,Q(x,t)]$
and $M$-operator with $P=1$, $f_{P}=1$, $V_{0}(x,t)=[I,Q(x,t)]$ and $%
V(x,t,\lambda )=[I,Q(x,t)]-\lambda I$.

\n
\textbf{Example 1.}
Consider algebra $\bf{g}\simeq \mathbf{C}_{2}$. The positive roots are $%
e_{1}\pm e_{2}$, $2e_{1}$ and $2e_{2}$ and thus: 
\begin{eqnarray*}
Q &=&Q_{1\overline{2}%
}E_{e_{1}-e_{2}}+Q_{12}E_{e_{1}+e_{2}}+Q_{11}E_{2e_{1}}+Q_{22}E_{2e_{2}} \\
&&+Q_{\overline{1}2}E_{-e_{1}+e_{2}}+Q_{\overline{1}\overline{2}%
}E_{-e_{1}-e_{2}}+Q_{\overline{1}\overline{1}}E_{-2e_{1}}+Q_{\overline{2}%
\overline{2}}E_{-2e_{2}}
\end{eqnarray*}
The system of NLEE for the components of $Q$ is 
\begin{eqnarray}
i(J_{1}-J_{2})Q_{1\overline{2},t}-i(I_{1}-I_{2})Q_{1\overline{2},x}+2\kappa
(Q_{\overline{2}\overline{2}}Q_{12}-Q_{11}Q_{\overline{1}\overline{2}}) &=&0
\nonumber \\
i(J_{1}+J_{2})Q_{12,t}-i(I_{1}+I_{2})Q_{12,x}-2\kappa (Q_{22}Q_{1\overline{2}%
}+Q_{11}Q_{\overline{1}2}) &=&0  \nonumber \\
iJ_{1}Q_{11,t}-iI_{1}Q_{11,x}+2\kappa Q_{1\overline{2}}Q_{12} &=&0  \nonumber
\\
iJ_{2}Q_{22,t}-iI_{2}Q_{22,x}+2\kappa Q_{\overline{1}2}Q_{12} &=&0  \nonumber
\\
i(J_{2}-J_{1})Q_{\overline{1}2,t}-i(I_{2}-I_{1})Q_{\overline{1}2,x}+2\kappa
(Q_{\overline{1}\overline{1}}Q_{12}-Q_{22}Q_{\overline{1}\overline{2}}) &=&0
\nonumber \\
-i(J_{1}+J_{2})Q_{\overline{1}\overline{2},t}+i(I_{1}+I_{2})Q_{\overline{1}%
\overline{2},x}+2\kappa (Q_{\overline{1}\overline{1}}Q_{1\overline{2}}+Q_{%
\overline{2}\overline{2}}Q_{\overline{1}2}) &=&0  \nonumber \\
-iJ_{1}Q_{\overline{1}\overline{1},t}+iI_{1}Q_{\overline{1}\overline{1}%
,x}-2\kappa Q_{\overline{1}\overline{2}}Q_{\overline{1}2} &=&0  \nonumber \\
-iJ_{2}Q_{\overline{2}\overline{2},t}+iI_{2}Q_{\overline{2}\overline{2}%
,x}-2\kappa Q_{\overline{1}\overline{2}}Q_{1\overline{2}} &=&0
\label{N wave system}
\end{eqnarray}
where $\kappa =J_{1}I_{2}-J_{2}I_{1}$.$\qed $ \\
The NLS-type equation \cite{VG*86}
\begin{equation}
iq_{t}+\text{\textrm{ad}}_{J}^{-1}q_{xx}-iP_{0}[q,\text{\textrm{ad}}%
_{J}^{-1}q_{x}]+\frac{1}{2}[q,(\mathbf{1}-P_{0})[q,\text{\textrm{ad}}%
_{J}^{-1}q]]=0,  \label{NLS}
\end{equation}
can be obtained, using $V(x,t,\lambda )=V_{0}+\lambda V_{1}-\lambda ^{2}J$.
When $J$ is a regular element, \textrm{ad}$_{J}$ is an invertible operator.
Here $P_{0}=$\textrm{ad}$_{J}^{-1}.$\textrm{ad}$_{J}$ is the projector on $%
\bf{g}\mathbf{/}\bf{h}$. From (\ref{compatibility}) we have the system

\begin{eqnarray*}
i{\frac{dV_{0}}{dx}-}i{\frac{dq}{dt}+}[q,V_{0}] &=&0 \\
i{\frac{dV_{1}}{dx}+[q,}V_{1}]-[J,V_{0}] &=&0 \\
\lbrack V_{1}-q,J] &=&0
\end{eqnarray*}
which can be resolved to give (\ref{NLS}), see \cite{VG*86}. The NLS
equations on symmetric spaces (where $J=\left( 
\begin{array}{cc}
\mathbf{1} & 0 \\ 
0 & \mathbf{-1}
\end{array}
\right) $) are considered in \cite{ForKu}.

\subsection{Dressing procedure}

The main goal of the dressing method is, starting from a FAS $\chi _{0}^{\pm
}(x,\lambda )$ of a Lax operator $L_{0}$ with a potential $q_{(0)}$ to construct a new
singular solution $\chi _{1}^{\pm }(x,\lambda )$ of (\ref{ZS}) with
singularities located at prescribed positions $\lambda _{1}^{\pm }$. The new
solutions $\chi _{1}^{\pm }(x,\lambda )$ will correspond to a new potential,
say $q_{(1)}$ of $L_{1}$, with two additional discrete eigenvalues $\lambda
_{1}^{\pm }$. \\
The new solution is related to the initial one by a dressing factor $%
u(x,\lambda )$: 
\[
\chi _{1}^{\pm }(x,\lambda )=u(x,\lambda )\chi _{0}^{\pm }(x,\lambda
)u_{-}^{-1}(\lambda ),\qquad u_{-}(\lambda )=\lim_{x\to -\infty }u(x,\lambda
). 
\]
Then $u(x,\lambda )$ obviously must satisfy the equation 
\begin{equation}
i{\frac{du}{dx}}+q_{(1)}(x)u(x,\lambda )-u(x,\lambda )q_{(0)}(x)-\lambda
[J,u(x,\lambda )]=0,  \label{diff u}
\end{equation}
and the normalization condition $\lim_{\lambda \to \infty }u(x,\lambda )=%
\mathbf{1}$. $\chi _{i}^{\pm }(x,\lambda )$, $i=0,1$ and $u(x,\lambda )$
must belong to the corresponding group $\bf{G}$. By construction $%
u(x,\lambda )$ has poles or/and zeroes at $\lambda _{1}^{\pm }$. All
quantities bellow, related to $L_{i}$ (\ref{ZS}) with potential $q_{(i)}(x)$
will be supplied with the corresponding index $i$. Their scattering data are
related by: 
\begin{eqnarray}
&&S_{(1)}^{\pm }(\lambda )=u_{-}(\lambda )S_{(0)}^{\pm }(\lambda
)u_{-}^{-1}(\lambda ),\qquad T_{(1)}^{\pm }(\lambda )=u_{+}(\lambda
)T_{(0)}^{\pm }(\lambda )u_{+}^{-1}(\lambda ),  \label{eq:SD1-0} \\
&&D_{(1)}^{\pm }(\lambda )=u_{+}(\lambda )D_{(0)}^{\pm }(\lambda
)u_{-}^{-1}(\lambda ),\qquad u_{\pm }(\lambda )=\lim_{x\to \pm \infty
}u(x,\lambda ).  \label{uDu}
\end{eqnarray}
Since the limits $u_{\pm }(\lambda )$ are $x$-independent and belong to the
Cartan subgroup $\bf{H}$ of $\bf{G}$, so $S_{(1)}^{\pm }(\lambda )$, $%
T_{(1)}^{\pm }(\lambda )$ are of the form (\ref{GaussTSD}). \\
If (\ref{diff u}) is satisfied then one can see that $L_{1}=uL_{0}u^{-1}$
and therefore $[L_{1},M_{1}]=0$ where
\begin{equation}
M_{1}=uM_{0}u^{-1}  \label{dressed M}
\end{equation}
The dressed potential $q_{(1)}(x,t)$ satisfies the same NLEE as $%
q_{(0)}(x,t) $ since $M_{1}$ (\ref{dressed M}) has a potential in the same
polynomial form in $\lambda $ as in $M_{0}$ due to the fact that the dressed
FAS are solutions to a RH problem of the same type (\ref{RH probl}) for all
values of the additional parameter $t$. \\
The simplest case $\bf{g}\simeq \mathbf{A}_{r}$ (or, rather $gl(r+1)$ ) is
solved in the classical papers \cite{ZMNP,Za*Sh}. The dressing factor is 
\begin{equation}
u(x,\lambda )=\mathbf{1}+{\frac{\lambda _{1}^{-}-\lambda _{1}^{+}}{\lambda
-\lambda _{1}^{-}}}P(x),  \label{sl(N) dressing }
\end{equation}
where the projector $P(x)$ can be chosen in the form: 
\[
P(x)={\frac{|n(x)\rangle \langle m(x)|}{\langle m(x)|n(x)\rangle }} 
\]
with $|n(x)\rangle =\chi _{0}^{+}(x,\lambda _{1}^{+})|n_{0}\rangle $, $%
\langle m(x)|=\langle m_{0}|\hat{\chi}_{0}^{-}(x,\lambda _{1}^{-})$; $%
|n_{0}\rangle $, $|m_{0}\rangle $ are constant vector-columns ( $\langle
m_{0}|=(|m_{0}\rangle )^{T}$). It can be easily checked that 
\[
q_{(1)}(x)=q_{(0)}(x)+(\lambda _{1}^{-}-\lambda _{1}^{+})[J,P(x)] 
\]
In fact $u(x,\lambda )$ (\ref{sl(N) dressing }) belongs not to $SL(r+1)$,
but to $GL(r+1)$. $\det (u)$ depends only on $\lambda $ and it is not a
problem to multiply $u(x,\lambda )$ by an appropriate scalar and thus to
adjust its determinant to 1. Such a multiplication easily goes through the
whole scheme outlined above.
We mention also the papers by Zakharov and Mikhailov \cite{Za*Mi} where they
generalized the dressing method and derived the soliton solutions for a
number of field theory models, related to the orthogonal and symplectic
algebras.

\section{Dressing factors related to the orthogonal and symplectic cases}

For the construction of the dressing factor for the case of orthogonal and
symplectic algebras we will assume that it contains singularities at $%
\lambda _{1}^{\pm }$ as proposed in \cite{VG*87}, \cite{Za*Mi}:

\begin{equation}
u(x,\lambda )=\mathbf{1}+(c_{\mu }(\lambda )-1)\pi _{1}(x)+(c_{\mu
}^{-1}(\lambda )-1)\pi _{-1}(x),\qquad  \label{u-anzatz}
\end{equation}
where for some constant $\mu $%
\begin{equation}
c_{\mu }(\lambda )=\left( {\frac{\lambda -\lambda _{1}^{+}}{\lambda -\lambda
_{1}^{-}}}\right) ^{\mu }.  \label{c-lambda}
\end{equation}
The two matrix-valued functions $\pi _{1}(x)$ and $\pi _{-1}(x)$ must
satisfy a system of algebraic equations ensuring that $u(x,\lambda )\in 
\bf{G}$, i.e. 
\begin{equation}
u^{-1}(x,\lambda )=Su^{T}(x,\lambda )S^{-1}  \label{def u }
\end{equation}
\smallskip where the matrix $S$ is
\begin{eqnarray}
S & = & \sum_{k=1}^{r}(-1)^{k+1}(E_{k\bar{k}}+E_{\bar{k}k})+(-1)^{r}E_{r+1,r+1},
\nonumber \\
&& \hskip60pt  \bar{k}=N+1-k,\qquad N=2r+1\qquad \bf{g}\simeq \mathbf{B}_{r},
\label{def S} \\
S & = & \sum_{k=1}^{r}(-1)^{k+1}(E_{k\bar{k}}-E_{\bar{k}k}), \nonumber \\
&& \hskip60pt  N=2r,\qquad 
\bar{k}=N+1-k,\qquad \bf{g}\simeq \mathbf{C}_{r},  \nonumber \\
S & = &\sum_{k=1}^{r}(-1)^{k+1}(E_{k\bar{k}}+E_{\bar{k}k}),\nonumber \\
&& \hskip60pt  N=2r,\qquad \bar{k}=N+1-k,\qquad \bf{g}\simeq \mathbf{D}_{r}.
\end{eqnarray}
Here $E_{kn}$ is an $N\times N$ matrix whose matrix elements are $%
(E_{kn})_{ij}=\delta _{ik}\delta _{nj}$ and $N$ is the dimension of the
typical representation of the corresponding algebra. We note also the
difference between the matrices $S$ for the symplectic and the orthogonal
case, namely: 
\begin{eqnarray*}
S^{-1} &=&S^{T}=S\text{ for the orthogonal  groups (algebras) and} \\
S^{-1} &=&S^{T}=-S\text{ for the symplectic groups (algebras).}
\end{eqnarray*}
The algebraic equations, following from the condition that (\ref{def u })
or, equivalently, 
\[
u(x,\lambda )Su^{T}(x,\lambda )S^{-1}\equiv \mathbf{1} 
\]
should hold identically with respect to $\lambda $ are:
\begin{center}
\begin{eqnarray}
\pi _{1}S\pi _{1}^{T}S^{-1} =\pi _{-1}S\pi _{-1}^{T}S^{-1}=0,
\label{pi S pi} \\
\pi _{1}+S\pi _{1}^{T}S^{-1}-\pi _{1}S\pi _{-1}^{T}S^{-1}-\pi _{-1}S\pi
_{1}^{T}S^{-1} =0,  \label{alg pi} \\
\pi _{-1}+S\pi _{-1}^{T}S^{-1}-\pi _{1}S\pi _{-1}^{T}S^{-1}-\pi _{-1}S\pi
_{1}^{T}S^{-1} =0  \label{alg pi 2}
\end{eqnarray}
\end{center}
The equations for $\pi _{\pm 1}(x)$ following from (\ref{diff u}), (\ref
{u-anzatz}) keeping in mind that it should also hold identically with
respect to $\lambda $ (i.e. when $\lambda \rightarrow \lambda _{1}^{\pm }$
and $\lambda \rightarrow \infty $) are:
\begin{eqnarray}
i{\frac{d\pi _{1}(x)}{dx}}+q_{(1)}(x)\pi _{1}(x)-\pi
_{1}(x)q_{(0)}(x)-\lambda _{1}^{-}[J,\pi _{1}(x)] &=&0,  \label{diff pi 11}
\\
i{\frac{d\pi _{-1}(x)}{dx}}+q_{(1)}(x)\pi _{-1}(x)-\pi
_{-1}(x)q_{(0)}(x)-\lambda _{1}^{+}[J,\pi _{-1}(x)] &=&0,  \label{diff pi-12}
\end{eqnarray}
\begin{eqnarray}
q_{(1)}(x) &=&q_{(0)}(x)+\lim_{\lambda \rightarrow \infty }\lambda
[J,(c_{\mu }(\lambda )-1)\pi _{1}(x)+(c_{\mu }^{-1}(\lambda )-1)\pi _{-1}(x)]
\label{q-1} \\
&=&q_{(0)}(x)+\mu (\lambda _{1}^{-}-\lambda _{1}^{+})[J,\pi _{1}(x)-\pi
_{-1}(x)].
\end{eqnarray}
It is possible to find solutions of (\ref{pi S pi})-(\ref{alg pi 2}) of the
form
\begin{equation}
\pi _{1}=Ym^{T},\;\pi _{-1}=SXn^{T}S^{-1}.  \label{pi XY}
\end{equation}
\smallskip Here $X$, $Y$, $n$, $m$ are $N\times r_{1}$ rectangular matrices
where $r_{1}\leq N$ and $N$ is the dimension of the typical representation
of the corresponding algebra as in (\ref{def S}) . The system (\ref{alg pi}%
)-(\ref{alg pi 2}) can be rewritten as 
\begin{eqnarray*}
Ym^{T}(\mathbf{1}-nX^{T}) &=&-S(\mathbf{1}-Xn^{T})mY^{T}S^{-1} \\
Xn^{T}(\mathbf{1}-mY^{T}) &=&-S(\mathbf{1}-Ym^{T})nX^{T}S^{-1}
\end{eqnarray*}
and can be solved for $X$, $Y$ introducing two new matrices $A$ and $B$ (yet
arbitrary) \cite{Za*Mi} via the relations 
\[
XB^{T}=S(\mathbf{1}-Ym^{T})n,\;YA^{T}=S(\mathbf{1}-Xn^{T})m. 
\]
The solution with respect to $X$ and $Y$ of the algebraic equations arising
from (\ref{pi S pi})-(\ref{alg pi 2}) can be written down in the form
\begin{eqnarray}
Y &=&(n+Sm(\rho ^{T})^{-1}B)R^{-1},\;\;X=(m+Sn\rho ^{-1}A)(R^{T})^{-1}
\label{XY solved} \\
\rho &=&m^{T}n,\;\;R=\rho -\sigma A(\rho ^{T})^{-1}B  \nonumber
\end{eqnarray}
where the matrices $n$, $m$, $A$ and $B$ must satisfy
\begin{equation}
A^{T}=-\sigma A,\;\;B^{T}=-\sigma B,\;\;n^{T}Sn=m^{T}Sm=0
\label{alfa-beta constr}
\end{equation}
and $\sigma $ is a sign determined from $S^{-1}=\sigma S$, see (\ref{def S}). \\
For the orthogonal algebras $\sigma =1$. If we take $r_{1}=1$ then $A$, $B$
are $1\times 1$ matrices and according to (\ref{alfa-beta constr}) $%
A=B\equiv 0$. Hence $\pi _{\pm 1}$ (\ref{pi XY}) are projectors of rank $1$.
Taking 
\begin{equation}
m=S\chi _{0}^{-}(x,t,\lambda _{1}^{-})S^{-1}|m_{0}\rangle ,\;n=\chi
_{0}^{+}(x,t,\lambda _{1}^{+})|n_{0}\rangle  \label{mn vectors}
\end{equation}
where $|n_{0}\rangle $, $|m_{0}\rangle $ are constant vector-columns ( $%
\langle m_{0}|=(|m_{0}\rangle )^{T}$) we have the following result \cite
{GGIK} which we quote for completeness:
\begin{prop}
Let $\bf{g}\simeq \mathbf{B}_{r}\ $or $\mathbf{D}_{r}$, $\chi _{0}^{\pm
}(x,t,\lambda )$ - the FAS for the Lax-pair $L_{0}$, $M_{0}$ and $\mu =1$ in
(\ref{c-lambda}). Then $\pi _{\pm 1}$ (\ref{u-anzatz}) and the Lax potential
have the form 
\begin{eqnarray}
\pi _{1}(x,t) & = & \nonumber \\
& & \hskip-21pt \chi _{0}^{+}(x,t,\lambda _{1}^{+})|n_{0}\rangle \left(
\langle m_{0}|\widehat{\chi }_{0}^{-}(x,t,\lambda _{1}^{-})\chi
_{0}^{+}(x,t,\lambda _{1}^{+})|n_{0}\rangle \right) ^{-1}\langle m_{0}|%
\widehat{\chi }_{0}^{-}(x,t,\lambda _{1}^{-}),  \nonumber \\
\pi _{-1}(x,t) & = & S \pi_{1}^{T}S^{-1},  \nonumber \\
q_{(1)}(x,t) & = & q_{(0)}(x,t)+(\lambda _{1}^{-}-\lambda _{1}^{+})[J,\pi
_{1}-\pi _{-1}]  \label{Br-Dr},
\end{eqnarray}
where $|n_{0}\rangle $, $|m_{0}\rangle $ are constant vectors from the
typical representation, such that $\langle m_{0}|S|m_{0}\rangle =\langle
n_{0}|S|n_{0}\rangle =0$.
\end{prop}
\begin{rmk}
$\pi _{\pm 1}$ are mutually orthogonal projectors, $\pi _{1}-\pi _{-1}\in 
\bf{g}$ and one can see that $u(x,t,\lambda )=\exp \left( (\ln
c_{1}(\lambda ))(\pi _{1}(x,t)-\pi _{-1}(x,t))\right) \in \bf{G}$.
\end{rmk}
\textbf{Proof}: The proof is based on a direct verification of (\ref{diff pi
11}), (\ref{diff pi-12}), see \cite{GGIK} for the details and examples,
related to the N-wave type equation (\ref{N wave}).$\qed $ \\
When we consider $\bf{g}\simeq \mathbf{C}_{r}$ however, even when $r_{1}=1$%
, $A$ and $B$ are in general not equal to zero. We state the result in the
following proposition:
\begin{prop}
\label{prop2}
Let $\bf{g}\simeq \mathbf{C}_{r}$, $r_{1}=1$, $\chi _{0}^{\pm
}(x,t,\lambda )$ - the FAS for the Lax pair $L_{0}$, $M_{0}$ and $\mu =1$ in
(\ref{c-lambda}). Then $\pi _{\pm 1}$ (\ref{u-anzatz}) and the Lax potential
have the form 
\begin{eqnarray}
\label{haha}
\pi _{1}(x,t) & = & \frac{(\rho n+BSm)m^{T}}{\rho ^{2}+AB}, \\
\pi _{-1}(x,t) & = & \frac{S(\rho m+ASn)n^{T}S^{-1}}{\rho ^{2}+AB},  \label{Pi-A-B} \\
q_{(1)}(x,t) &=&q_{(0)}(x,t)+(\lambda _{1}^{-}-\lambda _{1}^{+})[J,\pi
_{1}-\pi _{-1}]  \label{q for Cr}
\end{eqnarray}
with 
\begin{eqnarray}
n(x,t) & = & \chi_{0}^{+}(x,t,\lambda _{1}^{+})|n_{0}\rangle, \nonumber \\ 
m^{T}(x,t) & = & \langle m_{0}|\widehat{\chi }_{0}^{-}(x,t,\lambda _{1}^{-}), \nonumber \\
\rho(x,t) & = & m^{T}(x,t)n(x,t), \nonumber \\
\label{39}
A(x,t) & = & -(\lambda _{1}^{-}-\lambda _{1}^{+})\langle m_{0}|\widehat{\chi }%
_{0}^{-}(x,t,\lambda _{1}^{-})\stackrel{.}{\chi }_{0}^{-}(x,t,\lambda
_{1}^{-})S|m_{0}\rangle ,  \\
\label{40}
B{}(x,t) & = &(\lambda _{1}^{-}-\lambda _{1}^{+})\langle n_{0}|S\widehat{%
\chi }_{0}^{+}(x,t,\lambda _{1}^{+})\stackrel{.}{\chi }_{0}^{+}(x,t,\lambda
_{1}^{+})|n_{0}\rangle ,  
\end{eqnarray}
where the dot denotes a derivative with respect to $\lambda $:  $\stackrel{.}{%
\chi }_{0}^{\pm }(x,t,\lambda )=\frac{\partial }{\partial \lambda }\chi
_{0}^{\pm }(x,t,\lambda )$.
\end{prop}
\textbf{Proof:} Substituting $\pi _{\pm 1}$ from (\ref{haha}) and (\ref{Pi-A-B}) into equations (%
\ref{diff pi 11})--(\ref{diff pi-12}) with $m$, $n$ and $\rho $ as above,
after some tedious calculations we receive that $A$ and $B$ must satisfy the
extra conditions
\begin{eqnarray*}
i{\frac{dA}{dx}} &=&-(\lambda _{1}^{-}-\lambda _{1}^{+})\langle m_{0}|%
\widehat{\chi }_{0}^{-}(\lambda _{1}^{-})J\chi _{0}^{-}(\lambda
_{1}^{-})S|m_{0}\rangle \; \\
i{\frac{dB}{dx}} &=&(\lambda _{1}^{-}-\lambda _{1}^{+})\langle n_{0}|S%
\widehat{\chi }_{0}^{+}(\lambda _{1}^{+})J\chi _{0}^{+}(\lambda
_{1}^{+})|n_{0}\rangle \;
\end{eqnarray*}
which can be resolved, using the fact that the FAS satisfy (\ref{ZS}). We
thus end up with expressions (\ref{39}) and (\ref{40}).$\qed $

\n
\textbf{Example 2.}
Consider $\bf{g}\simeq \mathbf{C}_{2}$. For the system (\ref{N wave system}%
) related to the $N$-wave equation (\ref{N wave}) $%
Q_{(1)}(x,t)=Q_{(0)}(x,t)+(\lambda _{1}^{-}-\lambda _{1}^{+})P_{0}(\pi
_{1}-\pi _{-1})$ (here we ignore the irrelevant diagonal part). If $%
Q_{(0)}(x,t)\equiv 0$ clearly 
\[
\chi _{0}^{\pm }(x,t,\lambda )=\exp \left( -i\lambda (Jx+It)\right) .
\]
Let us take the typical representation of $\mathbf{C}_{2}$ with a basis $%
|\gamma _{i}\rangle =|e_{i}\rangle $, $|\gamma _{\overline{i}}\rangle
=|-e_{i}\rangle $, $i=1,2$ and 
\begin{eqnarray*}
|n_{0}\rangle  &=&n_{01}|\gamma _{1}\rangle +n_{02}|\gamma _{2}\rangle +n_{0%
\overline{2}}|\gamma _{\overline{2}}\rangle +n_{0\overline{1}}|\gamma _{%
\overline{1}}\rangle  \\
|m_{0}\rangle  &=&m_{01}|\gamma _{1}\rangle +m_{02}|\gamma _{2}\rangle +m_{0%
\overline{2}}|\gamma _{\overline{2}}\rangle +m_{0\overline{1}}|\gamma _{%
\overline{1}}\rangle 
\end{eqnarray*}
with all constant parameters $n_{0i}$, $m_{0i}$ nonzero. Then from (\ref{q
for Cr}) we obtain the following solution 
\begin{eqnarray*}
Q_{1\overline{2}} &=&\frac{l}{\Delta}(n_{01}m_{02}\rho
e^{i\lambda _{1}^{-}z_{2}-i\lambda _{1}^{+}z_{1}}+n_{0\overline{2}}m_{0%
\overline{1}}\rho e^{-i\lambda _{1}^{-}z_{1}+i\lambda _{1}^{+}z_{2}} \\
&&-n_{01}n_{0\overline{2}}Ae^{-i\lambda _{1}^{+}(z_{1}-z_{2})}+m_{02}m_{0%
\overline{1}}Be^{-i\lambda _{1}^{-}(z_{1}-z_{2})}) \\
Q_{12} &=&\frac{l}{\Delta}(n_{01}m_{0\overline{2}}\rho
e^{-i\lambda _{1}^{-}z_{2}-i\lambda _{1}^{+}z_{1}}-n_{02}m_{0\overline{1}%
}\rho e^{-i\lambda _{1}^{-}z_{1}-i\lambda _{1}^{+}z_{2}} \\
&&+n_{01}n_{02}Ae^{-i\lambda _{1}^{+}(z_{1}+z_{2})}+m_{0\overline{2}}m_{0%
\overline{1}}Be^{-i\lambda _{1}^{-}(z_{1}+z_{2})})  \\
Q_{11} &=&\frac{l}{\sqrt{2}\Delta}(2n_{01}m_{0%
\overline{1}}\rho e^{-i(\lambda _{1}^{-}+\lambda
_{1}^{+})z_{1}}-n_{01}^{2}Ae^{-2i\lambda _{1}^{+}z_{1}}+m_{0\overline{1}%
}^{2}Be^{-2i\lambda _{1}^{-}z_{1}})  \\
Q_{22} &=&\frac{l}{\sqrt{2}\Delta}(2n_{02}m_{0%
\overline{2}}\rho e^{-i(\lambda _{1}^{-}+\lambda
_{1}^{+})z_{2}}+n_{02}^{2}Ae^{-2i\lambda _{1}^{+}z_{2}}-m_{0\overline{2}%
}^{2}Be^{-2i\lambda _{1}^{-}z_{2}})  \\
Q_{\overline{1}2} &=&\frac{l}{\Delta}(n_{02}m_{01}\rho
e^{i\lambda _{1}^{-}z_{1}-i\lambda _{1}^{+}z_{2}}+n_{0\overline{1}}m_{0%
\overline{2}}\rho e^{-i\lambda _{1}^{-}z_{2}+i\lambda _{1}^{+}z_{1}} \\
&&+n_{02}n_{0\overline{1}}Ae^{i\lambda _{1}^{+}(z_{1}-z_{2})}-m_{01}m_{0%
\overline{2}}Be^{i\lambda _{1}^{-}(z_{1}-z_{2})})  \\
Q_{\overline{1}\overline{2}} &=&\frac{l}{\Delta}(n_{0%
\overline{2}}m_{01}\rho e^{i\lambda _{1}^{-}z_{1}+i\lambda
_{1}^{+}z_{2}}-n_{0\overline{1}}m_{02}\rho e^{i\lambda
_{1}^{-}z_{2}+i\lambda _{1}^{+}z_{1}} \\
&&+n_{0\overline{1}}n_{0\overline{2}}Ae^{i\lambda
_{1}^{+}(z_{1}+z_{2})}+m_{01}m_{02}Be^{i\lambda
_{1}^{-}(z_{1}+z_{2})})  \\
Q_{\overline{2}\overline{2}} &=&\frac{l}{\sqrt{2}\Delta}(2n_{0\overline{2}}m_{02}\rho e^{i(\lambda _{1}^{-}+\lambda
_{1}^{+})z_{2}}-n_{0\overline{2}}^{2}Ae^{2i\lambda
_{1}^{+}z_{2}}+m_{02}^{2}Be^{2i\lambda _{1}^{-}z_{2}})  \\
Q_{\overline{1}\overline{1}} &=&\frac{l}{\sqrt{2}\Delta}(2n_{0\overline{1}}m_{01}\rho e^{i(\lambda _{1}^{-}+\lambda
_{1}^{+})z_{1}}+n_{0\overline{1}}^{2}Ae^{2i\lambda
_{1}^{+}z_{1}}-m_{01}^{2}Be^{2i\lambda _{1}^{-}z_{1}}) 
\end{eqnarray*}
with the notations: 
\begin{eqnarray*}
l&=&\lambda _{1}^{-}-\lambda _{1}^{+}\\
A &=&2il(m_{01}m_{0\overline{1}%
}z_{1}-m_{02}m_{0\overline{2}}z_{2}),\text{ \ }B=2il(n_{01}n_{0\overline{1}}z_{1}-n_{02}n_{0\overline{2}}z_{2}) \\
\rho  &=&n_{01}m_{01}e^{ilz_{1}}+n_{02}m_{02}e^{ilz_{2}}+
n_{0\overline{2}}m_{0\overline{2}}e^{-ilz_{2}}+n_{0\overline{1}}m_{0\overline{1}%
}e^{-ilz_{1}}, \\
\Delta  &=&\rho ^{2}+AB, \\
z(x,t) &=&Jx+It,\;\;z_{k}(x,t)=J_{k}x+I_{k}t.
\end{eqnarray*}
The Cartan-Weyl generators are 
\begin{eqnarray*}
E_{e_{i}-e_{j}} &=&E_{ij}+(-1)^{i+j+1}E_{\overline{j}\overline{i}%
},\;\;E_{e_{i}+e_{j}}=E_{i\overline{j}}+(-1)^{i+j+1}E_{j\overline{i}%
},\;(i<j), \\
E_{2e_{k}} &=&\sqrt{2}E_{k\overline{k}},\;\;H_{e_{k}}=E_{kk}-E_{\overline{k}%
\overline{k}}.
\end{eqnarray*}
Note the terms $A$, $B$, linear with respect to $x$ and $t$ due to the
contributions from $\left( \frac{\partial }{\partial \lambda }\chi
_{0}^{\pm }(x,t,\lambda )\right) _{\lambda =\lambda _{1}^{\pm }}$. The
reduction 
\[
\lambda _{1}^{-}=\overline{\lambda _{1}^{+}},\;\;m_{0k}=\overline{n_{0k}}
\]
leads to finite solutions since in that case 
\[
\Delta =|\rho |^{2}+|A|^{2}\neq 0.
\]
In order to analyze the scattering data in this example we recall that our
regularity assumption $\alpha (J)>0$ for $\bf{g}\simeq \mathbf{C}_{2}$
implies $J_{1}>J_{2}>0$. Our initial solution is $Q_{(0)}(x,t)\equiv 0$ and $%
D_{(0)}^{\pm }(\lambda )=S_{(0)}^{\pm }(\lambda )=T_{(0)}^{\pm }(\lambda )=%
\mathbf{1}$. It is not difficult to calculate that 
\begin{eqnarray*}
\lim_{x\to -\infty }\pi _{1}(x,t) &=&|\gamma _{\overline{1}}\rangle \langle
\gamma _{\overline{1}}|\equiv E_{\overline{1}\overline{1}}\text{,\ }%
\lim_{x\to \infty }\pi _{1}(x,t)=|\gamma _{1}\rangle \langle \gamma
_{1}|\equiv E_{11} \\
\lim_{x\to -\infty }\pi _{-1}(x,t) &=&|\gamma _{1}\rangle \langle \gamma
_{1}|\text{,\ }\lim_{x\to \infty }\pi _{-1}(x,t)=|\gamma _{\overline{1}%
}\rangle \langle \gamma _{\overline{1}}|
\end{eqnarray*}
and thus from (\ref{uDu}) 
\begin{eqnarray*}
u_{\pm }(\lambda ) &=&\exp \left( \pm (\ln c_{1}(\lambda ))H_{e_{1}}\right) ,
\\
D_{(1)}^{\pm }(\lambda ) &=&\mathbf{1}.\exp \left( 2(\ln c_{1}(\lambda
))H_{e_{1}}\right) .
\end{eqnarray*}
Since for the $\mathbf{C}_{r}$ algebra $w_{0}=-\mathbf{1}$\textbf{, }$\omega
_{j}^{+}=e_{1}+...+e_{j}$ from (\ref{d_j}) we have 
\[
d_{(1)j}^{\pm }(\lambda )=d_{(0)j}^{\pm }(\lambda )\pm 2(\ln c_{1}(\lambda
))(e_{1},\omega _{j}^{+})=\pm 2\ln c_{1}(\lambda )\text{, \ }j=1,2.
\]
Since we obtained soliton solutions, corresponding to a reflectionless
potential, the scattering data on the continuous spectrum remained trivial (%
\ref{eq:SD1-0}): $S_{(1)}^{\pm }(\lambda )=T_{(1)}^{\pm }(\lambda )=\mathbf{1%
}$.$\qed $ \\
Next we examine one particular case, where $\pi _{1}+\pi _{-1}=\mathbf{1}$.
We start with the following Lemma.
\begin{lem}
\label{lemma_proj}
If $\pi _{1}+\pi _{-1}=\mathbf{1}$, the equations (\ref{pi S pi})---(\ref
{q-1}) are compatible if and only if
\end{lem}
\begin{enumerate}
\item  $\pi _{-1}=S\pi _{1}^{T}S^{-1}$
\item  $\pi _{\pm 1}$ are mutually orthogonal projectors ($\pi _{\pm
1}^{2}=\pi _{\pm 1}$, $\pi _{1}\pi _{-1}=\pi _{-1}\pi _{1}=0$)
\item  $\mu =\frac{1}{2}$
\end{enumerate}
\textbf{Proof}: Assume that (\ref{pi S pi}) holds. If $\pi _{1}+\pi _{-1}=%
\mathbf{1}$ then (\ref{alg pi}) is satisfied and from (\ref{alg pi 2}) we
have $\pi _{-1}=S\pi _{1}^{T}S^{-1}$. Then $\pi _{1}\pi _{-1}=\pi _{-1}\pi
_{1}=0$ and $\pi _{1}^{2}=\pi _{1}(1-\pi _{-1})=\pi _{1}$. From (\ref
{c-lambda}) and (\ref{q-1}) provided $\pi _{-1}=\mathbf{1-}\pi _{1}$ we have
\[
q_{(1)}(x)=q_{(0)}(x)+2\mu (\lambda _{1}^{-}-\lambda _{1}^{+})[J,\pi
_{1}(x)]. 
\]
On the other hand, summing up (\ref{diff pi 11}) and (\ref{diff pi-12}) we
have
\begin{equation}
q_{(1)}(x)=q_{(0)}(x)+(\lambda _{1}^{-}-\lambda _{1}^{+})[J,\pi _{1}(x)]
\label{new q}
\end{equation}
and therefore $\mu =\frac{1}{2}$.$\qed $ \\
Clearly $c_{\frac{1}{2}}(\lambda )$ is not a meromorphic function, but this
is not a real problem since in this case 
\begin{eqnarray*}
u(x,\lambda ) &=&c_{\frac{1}{2}}^{-1}(\lambda )\left( \mathbf{1}+(c_{\frac{1%
}{2}}^{2}(\lambda )-1)\pi _{1}(x)\right) , \\
u^{-1}(x,\lambda ) &=&c_{\frac{1}{2}}(\lambda )\left( \mathbf{1}+\left( c_{\frac{1}{2}%
}^{-2}(\lambda )-1\right) \pi _{1}(x)\right)
\end{eqnarray*}
and up to the irrelevant scalar multiplier $c_{\frac{1}{2}}^{-1}(\lambda )$
the dressing factor is a meromorphic function of $\lambda $.
\begin{prop}
If $\chi _{0}^{\pm }(x,t,\lambda )$ are the FAS for $L_{0}$ and $M_{0}$ the
projector $\pi _{1}$ of rank $r$ (in the context of Lemma \ref{lemma_proj}) has the form 
\begin{equation}
\pi _{1}(x,t)=\sum_{i,k=1}^{r}\chi _{0}^{+}(x,t,\lambda
_{1}^{+})|n_{0}^{i}\rangle R_{ik}^{-1}\langle m_{0}^{k}|\widehat{\chi }%
_{0}^{-}(x,t,\lambda _{1}^{-})  \label{projector rk r}
\end{equation}
where $|n_{0}^{i}\rangle $, $|m_{0}^{i}\rangle $ are constant vectors from
the corresponding (typical) representation,
\[
R_{ik}(x,t)=\langle m_{0}^{i}|\widehat{\chi }_{0}^{-}(x,t,\lambda
_{1}^{-})\chi _{0}^{+}(x,t,\lambda _{1}^{+})|n_{0}^{k}\rangle ,
\]
provided $\det (R)\neq 0$, $\langle m_{0}^{i}|S|m_{0}^{k}\rangle =\langle
n_{0}^{i}|S|n_{0}^{k}\rangle =0$.
\end{prop}
\begin{rmk}
Since rank$(\pi _{1})=$rank$(\pi _{-1})=r$ and $\pi _{1}+\pi _{-1}=\mathbf{1}
$ the proposed construction works only for algebras with $2r$ dimensional
typical representations, i.e. $\bf{g}\simeq \mathbf{C}_{r}$, $\mathbf{D}%
_{r}$.
\end{rmk}
\textbf{Proof}: $\pi _{-1}=S\pi _{1}^{T}S^{-1}=\sum_{i,k=1}^{r}\chi
_{0}^{-}(x,t,\lambda _{1}^{-})S|m_{0i}\rangle \widehat{R}_{ik}^{T}\langle
n_{0k}|S^{-1}\widehat{\chi }_{0}^{+}(x,t,\lambda _{1}^{+})$ and the
condition $\langle m_{0i}|S|m_{0k}\rangle =\langle n_{0i}|S|n_{0k}\rangle =0$
guarantees that $\pi _{1}\pi _{-1}=\pi _{-1}\pi _{1}=0$. Since the
projectors $\pi _{1}$ and $\pi _{-1}$ commute, they can be diagonalised
simultaneously, and $\pi _{1}$ has an eigenvalue $1$ at the places where $%
\pi _{-1}$ has an eigenvalue $0$ and vice-versa. It means that $\pi _{1}+\pi
_{-1}=\mathbf{1}$ if the representation space is $2r$ dimensional.
Furthermore one can easily verify that $\pi _{1}$ satisfies the equation
\[
i{\frac{d\pi _{1}}{dx}}+[q_{(0)},\pi _{1}]+\lambda _{1}^{-}\pi _{1}J-\lambda
_{1}^{+}J\pi _{1}-(\lambda _{1}^{-}-\lambda _{1}^{+})\pi _{1}J\pi _{1}=0 
\]
which is equivalent to (\ref{diff pi 11}) in the case $\pi _{1}+\pi _{-1}=%
\mathbf{1}$, taking into account (\ref{new q}).$\qed $ \\
\textbf{Example 3.}
For the $N$-wave equation (\ref{N wave}) and $\bf{g}\simeq \mathbf{C}_{2}$
i.e. for the system (\ref{N wave system}) with $Q_{(0)}(x,t)\equiv 0$, if we
take the typical representation of $\mathbf{C}_{2}$ and 
\begin{eqnarray*}
|n_{0}^{1}\rangle  &=&|\gamma _{1}\rangle +a|\gamma _{2}\rangle +b|\gamma _{%
\overline{2}}\rangle -ab|\gamma _{\overline{1}}\rangle , \\
|n_{0}^{2}\rangle  &=&|\gamma _{1}\rangle -a|\gamma _{2}\rangle +c|\gamma _{%
\overline{2}}\rangle +ac|\gamma _{\overline{1}}\rangle , \\
|m_{0}^{1}\rangle  &=&|n_{0}^{2}\rangle ,\;|m_{0}^{2}\rangle
=|n_{0}^{1}\rangle 
\end{eqnarray*}
with real positive constant parameters $a$, $b$ and $c$, the construction (%
\ref{projector rk r}) with the reduction 
\begin{equation}
\lambda _{1}^{-}=\overline{\lambda _{1}^{+}}  \label{lambda inv}
\end{equation}
gives the following solution: 
\begin{eqnarray*}
Q_{1\overline{2}} &=&4i\nu _{1}a^{2}(b-c)(b+c)e^{-i\mu
_{1}(z_{1}-z_{2})}\{\cosh [\nu _{1}(z_{1}-z_{2})-\ln a]\}/\Delta  \\
Q_{12} &=&-8i\nu _{1}a^{2}\sqrt{bc}(b+c)e^{-i\mu _{1}(z_{1}+z_{2})}\{\cosh
[\nu _{1}(z_{1}+z_{2})-\ln \sqrt{bc}]\}/\Delta  \\
Q_{11} &=&4\sqrt{2}i\nu _{1}a^{2}\sqrt{bc}(b-c)e^{-2i\mu _{1}z_{1}}\{\sinh
[2\nu _{1}z_{2}+\ln \frac{a}{\sqrt{bc}}]\}/\Delta  \\
Q_{22} &=&-4\sqrt{2}i\nu _{1}a^{2}\sqrt{bc}(b-c)e^{-2i\mu _{1}z_{2}}\{\sinh
[2\nu _{1}z_{1}-\ln a\sqrt{bc}]\}/\Delta  \\
Q_{\overline{1}2} &=&4i\nu _{1}a^{2}(b-c)(b+c)e^{i\mu
_{1}(z_{1}-z_{2})}\{\cosh [\nu _{1}(z_{1}-z_{2})-\ln a]\}/\Delta  \\
Q_{\overline{1}\overline{2}} &=&-8i\nu _{1}a^{2}\sqrt{bc}(b+c)e^{i\mu
_{1}(z_{1}+z_{2})}\{\cosh [\nu _{1}(z_{1}+z_{2})-\ln \sqrt{bc}]\}/\Delta  \\
Q_{\overline{2}\overline{2}} &=&-4\sqrt{2}i\nu _{1}a^{2}(b-c)e^{2i\mu
_{1}z_{2}}\{\cosh [2\nu _{1}z_{1}-\ln a\sqrt{bc}]\}/\Delta  \\
Q_{\overline{1}\overline{1}} &=&4\sqrt{2}i\nu _{1}a(b-c)e^{2i\mu _{1}z_{1}}a%
\sqrt{bc}\{\cosh [2\nu _{1}z_{2}+\ln \frac{a}{\sqrt{bc}}]\}/\Delta 
\end{eqnarray*}
The notations are as follows: 
\begin{eqnarray}
\label{44}
\nu _{1} & = & \mbox{Im } (\lambda _{1}^{+}) > 0,\qquad \mu _{1} = \mbox{Re } (\lambda _{1}^{+}) \\
\label{lambda}
\Delta (x,t) &=&a^{2}(b+c)^{2}+a^{2}(b-c)^{2}\cosh [2\nu
_{1}(z_{1}-z_{2})-\ln a^{2}] \nonumber \\
& & + 4a^{2}\cosh [2\nu _{1}(z_{1}+z_{2})-\ln bc], \nonumber \\
\Delta (x,t) &\neq &0; \nonumber \\
z(x,t) & = & Jx+It,\text{ }z_{k}(x,t)=J_{k}x+I_{k}t. \nonumber
\end{eqnarray}
In order to analyze the scattering data in this example we recall that $%
J_{1}>J_{2}>0$. It is not difficult to calculate that 
\[
\lim_{x\to -\infty }\pi _{1}(x,t)=E_{\overline{1}\overline{1}}+E_{\overline{2%
}\overline{2}}\text{,\ }\lim_{x\to \infty }\pi _{1}(x,t)=E_{11}+E_{22}
\]
and thus 
\begin{eqnarray*}
u_{\pm }(\lambda ) &=&\exp \left( \pm (\ln c_{1}(\lambda
))(H_{e_{1}}+H_{e_{2}})\right) \text{, } \\
D_{(1)}^{\pm }(\lambda ) &=&\exp \left( 2(\ln c_{1}(\lambda
))(H_{e_{1}}+H_{e_{2}})\right) .
\end{eqnarray*}
From (\ref{d_j}) we have 
\begin{eqnarray*}
d_{(1)j}^{\pm }(\lambda ) &=&d_{(0)j}^{\pm }(\lambda )\pm 2(\ln
c_{1}(\lambda ))(\omega _{j}^{+},e_{1}+e_{2}) \\
d_{(1)1}^{\pm }(\lambda ) &=&\pm 2\ln c_{1}(\lambda )\text{,\ }d_{(1)2}^{\pm
}(\lambda )=\pm 4\ln c_{1}(\lambda ).
\end{eqnarray*}
$\qed $ \\
Examples for the NLS-type equation (\ref{NLS}) can be easily constructed,
using the FAS $\chi _{0}^{\pm }(x,t,\lambda )=\exp \left( -i\lambda
J(x+\lambda t)\right) $ when $q_{(0)}(x,t)\equiv 0$ .

\section{Generating solutions for systems, related to \\ subalgebras}

\n
As a byproduct of the presented general constructions for the orthogonal and
symplectic algebras, we can generate solutions for their subalgebras \cite{GGIK}. As an
example we consider some $sl(2)$ solutions for the NLS-type equation (\ref
{NLS}) and their relation to the dressing construction for the $sl(N)$
algebra (\ref{sl(N) dressing }). If the dressed solution has the form 
\[
q_{(1)}(x,t)=q(x,t)E_{\alpha }+\widetilde{q}(x,t)E_{-\alpha }
\]
then $q$ and $\widetilde{q}$ must satisfy the equation (\ref{NLS}) related to $sl(2)$%
, which is of the form 
\begin{eqnarray}
iq_{t}+\omega _{1}q_{xx}+\omega _{2}q^{2}\widetilde{q} &=&0 \nonumber \\
\label{NLS for sl(2)} 
i\widetilde{q}_{t}-\omega _{1}\widetilde{q}_{xx}-\omega _{2}q\widetilde{q}%
^{2} &=&0  
\end{eqnarray}
with some constant coefficients $\omega _{1}$, $\omega _{2}$ depending on the
length of the root $\alpha $. \\
\textbf{Example 4.}
For the NLS-type equation (\ref{NLS}) when $q_{(0)}(x,t)\equiv 0$,  clearly \linebreak 
$q_{(1)}(x,t)=l[J,\pi _{1}(x,t)-\pi
_{-1}(x,t)]$ and $\chi _{0}^{\pm }(x,t,\lambda )=\exp \left( -i\lambda
J(x+\lambda t)\right) $. Let us take $\bf{g}\simeq \mathbf{C}_{r}$ ( $%
\mathbf{C}_{2}$ is sufficient for what follows) with its typical
representation. Let the basis in this representation be $|\gamma _{k}\rangle
=|e_{k}\rangle $, $|\gamma _{\overline{k}}\rangle =|-e_{k}\rangle $, $%
k=1,2,...,r$ and 
\begin{eqnarray*}
|n_{0}\rangle  &=&n_{01}|\gamma _{1}\rangle +n_{0\overline{1}}|\gamma _{%
\overline{1}}\rangle , \\
|m_{0}\rangle  &=&m_{01}|\gamma _{1}\rangle +m_{0\overline{1}}|\gamma _{%
\overline{1}}\rangle 
\end{eqnarray*}
with all constant parameters $n_{0i}$, $m_{0i}$ nonzero. From Proposition \ref{prop2} we have 
\[
q_{(1)}(x,t)=q(x,t)E_{2e_{1}}+\widetilde{q}(x,t)E_{-2e_{1}}
\]
where 
\begin{eqnarray}
q(x,t) &=&\frac{\eta J_{1}l}{\sqrt{2}\Delta}[(1-f^{-})e^{-2Z^{+}}+\eta \nu (1+f^{+})e^{-2Z^{-}}]  
\nonumber \\
\widetilde{q}(x,t) &=&-\frac{\nu J_{1}l}{\sqrt{2}\Delta}[(1-f^{+})e^{2Z^{-}}+\eta \nu (1+f^{-})e^{2Z^{+}}] 
\label{sol sl2}
\end{eqnarray}
with the following notations: 
\begin{eqnarray*}
l&=&\lambda _{1}^{-}-\lambda _{1}^{+}, \\
\Delta (x,t) &=&\left( e^{(Z^{-}-Z^{+})}+\eta \nu
e^{-(Z^{-}-Z^{+})}\right) ^{2}+4\eta \nu f^{+}f^{-}, \\
Z^{\pm }(x,t) &=&iJ_{1}\lambda _{1}^{\pm }(x+\lambda _{1}^{\pm }t), \\
f^{\pm }(x,t) &=&iJ_{1}l(x+2\lambda
_{1}^{\pm }t),
\end{eqnarray*}
$\eta =\frac{m_{0\overline{1}}}{m_{01}}$, $\nu =\frac{n_{0\overline{1}}%
}{n_{01}}$ are constants. For example, under the involution (\ref{lambda inv}%
), (\ref{44}) and $\nu =\overline{\eta }$, 
\[
\Delta (x,t)=4|\eta |^{2}\{\cosh ^{2}[2J_{1}\nu _{1}(x+2\mu _{1}t)-\ln
|\eta |]+4J_{1}^{2}\nu _{1}^{2}[(x+2\mu _{1}t)^{2}+4\nu
_{1}^{2}t^{2}]\}\neq 0,
\]
and the solutions are nonsingular. Note the linear terms with respect to $x$
and $t$ in $f^{\pm }$ due to the contributions from $\left( \frac{%
\partial }{\partial \lambda }\chi _{0}^{\pm }(x,t,\lambda )\right) _{\lambda
=\lambda _{1}^{\pm }}$. The solution (\ref{sol sl2}) of (\ref{NLS}) is also a solution of (%
\ref{NLS for sl(2)}) with coefficients: $\omega _{1}=(2J_{1})^{-1}$%
, $\omega _{2}=2/J_{1}$. The origin of this solution can be understood if we
apply two times the dressing construction for $sl(2)$ (\ref{sl(N) dressing }%
) with dressing factors 
\[
u_{2}(\lambda )=\left( \mathbf{1}+{\frac{\lambda _{2}^{-}-\lambda _{2}^{+}}{%
\lambda -\lambda _{2}^{-}}}P^{\prime }\right) \text{ \ }u_{1}(\lambda
)=\left( \mathbf{1}+{\frac{\lambda _{1}^{-}-\lambda _{1}^{+}}{\lambda
-\lambda _{1}^{-}}}P\right) 
\]
where: 
\begin{eqnarray*}
P &=&\frac{\chi _{0}^{+}(x,t,\lambda _{1}^{+})|n_{0}\rangle \langle m_{0}|%
\widehat{\chi }_{0}^{-}(x,t,\lambda _{1}^{-})}{\langle m_{0}|\widehat{\chi }%
_{0}^{-}(x,t,\lambda _{1}^{-})\chi _{0}^{+}(x,t,\lambda
_{1}^{+})|n_{0}\rangle } \\
P^{\prime } &=&\frac{u_{1}(\lambda _{2}^{+})\chi _{0}^{+}(x,t,\lambda
_{2}^{+})|n_{0}\rangle \langle m_{0}|\widehat{\chi }_{0}^{-}(x,t,\lambda
_{2}^{-})u_{1}^{-1}(\lambda _{2}^{-})}{\langle m_{0}|\widehat{\chi }
_{0}^{-}(x,t,\lambda _{2}^{-})u_{1}^{-1}(\lambda _{2}^{-})u_{1}(\lambda
_{2}^{+})\chi _{0}^{+}(x,t,\lambda _{2}^{+})|n_{0}\rangle } \\
q_{(2)} &=&[J,(\lambda _{2}^{-}-\lambda _{2}^{+})P^{\prime }+(\lambda
_{1}^{-}-\lambda _{1}^{+})P]
\end{eqnarray*}
and $\chi _{0}^{\pm }(x,t,\lambda )=\exp \left( -i\lambda J(x+\lambda
t)\right) $. Now $J=\mathrm{diag}(J_{1,}-J_{1})$. Note that we use the same
constant vectors $m_{0}=m_{01}|e_{1}\rangle +m_{02}|e_{2}\rangle $ and $%
n_{0}=n_{01}|e_{1}\rangle +n_{02}|e_{2}\rangle $ for both projectors. If we
take the limit $\lambda _{2}^{\pm }\rightarrow \lambda _{1}^{\pm }$, then $%
q_{(2)}=q^{+}E_{e_{1}-e_{2}}+q^{-}E_{-(e_{1}-e_{2})}$ and the solutions $%
q^{\pm }$ coincide with $q$ and $\widetilde{q}$ up to a constant factor $%
\sqrt{2}$ (it is because of the difference in the constant coefficients in
the systems of differential equations arising in both cases from (\ref{NLS}%
)) if we identify the constants $\eta =\frac{m_{02}}{m_{01}}$, $\nu =%
\frac{n_{02}}{n_{01}}$. It means that this solution represents a degenerate
two soliton solution, or a solution where the $sl(2)$-dressing factor has a
pole of order two. Solitons of the form (\ref{sol sl2}) for $sl(N)$ are also
examined in \cite{S}.$\qed $ \\
This simple example shows that the considered dressing constructions related
to $\mathbf{C}_{r}$ cannot be obtained by the simple-pole $sl(N)$ construction $ (\ref{sl(N) dressing })$ although 
$\mathbf{C}_{r}$ is a subalgebra of $sl(2r)$.

\section{Conclusions}

\n
We considered two constructions for the dressing factor (\ref{u-anzatz})
-one related to $\mathbf{C}_{r}$ (\ref{Pi-A-B}), (\ref{q for Cr}) and one
for $\mathbf{C}_{r}$,$\mathbf{D}_{r}$ (\ref{projector rk r}) and examples to
each one with $\mathbf{C}_{2}$. \\
It is known that the typical representation of $\mathbf{C}_{2}$ is
isomorphic to the spinor representation of $\mathbf{B}_{2}$. It is
interesting to be seen what type of solution for the spinor representation
of $\mathbf{B}_{2}$ corresponds to a solution of the form (\ref{q for Cr})
for the typical representation of $\mathbf{C}_{2}$. \\
Since the number of NLEE arising in a system is big, different reductions on
the such constructed solutions could be imposed \cite{2}. Some examples of
reduced $N$-wave equations are given in \cite{GGIK}, \cite{GGK}. Also examples
with potentials from the real forms of the algebras or symmetric spaces \cite
{ForKu} can be constructed.

\section{Acknowledgments}

The author is indebted to Prof. V.S. Gerdjikov for an introduction to the
problem treated in this paper and for many valuable discussions.

\end{document}